# A Triangulated Model to Assess Adoption of Virtual Learning Environments in Primary Schools

Elena Codreanu[1,2,3,4], Christine Michel[1,3], Marc-Eric Bobillier-Chaumon[1,2] and Olivier Vigneau[4]

[1] Université de Lyon, 92, Rue Pasteur, 69007, Lyon, France
[2] Université Lyon2, GRePS, EA 4163, 5, Avenue Pierre Mendès France, 69676, Bron, France
[3] INSA-Lyon, LIRIS, UMR5205, F-69621, 20, Avenue Albert Einstein, 69621, Villeurbanne Cedex, France
[4] WebServices pour l'Education, 22, Rue Legendre, 75017, Paris, France
{elena.codreanu, marc-eric.bobillier-chaumon}@univ-lyon2.fr, christine.michel@insa-lyon.fr,
olivier.vgneau@web-education.net

Abstract: The objective of this paper is to highlight the existing theoretical approaches which study the issue of technological adoption, and to establish a triangulated model to explore Virtual Learning Environments adoption in primary schools. The theoretical models cover three approaches: the social acceptance, the practical acceptance and the situated acceptance. Our triangulated model proposes to explore three types of factors: technological factors, activity and task factors and perceptual factors in order to assess technological adoption.

## 1 INTRODUCTION

When new technology is deployed in schools, it is generally expected to improve educational practices overall. New technologies are associated to quick change, modernisation, and improved efficiency. These resonate with contemporary issues in education like innovation, modernisation and democratisation of schools. This political will of modernisation remains quite general and it is not backed up by scientific information on various situations and use contexts. Such research is increasingly necessary as existing studies (Blin and Munro, 2008, Cuban et al., 2001, Jonsson, 2007) highlighted low use of technologies available in schools. Also, contemporary technologies become more complex, flexible and interconnected. VLEs (Virtual Learning Environments) are a typical example of complex technology, with services designed for teaching, learning, school management; addressed to different public: teachers, parents, students, and available in various contexts: at home, at school and in mobility situation. The term VLE has different connotations from country to country. In UK, VLEs were designed primarily as collaborative learning spaces to which administrative modules were later added. In this view, a VLE is *"learner centred and facilitates the offering of active learning opportunities, including specific tutor guidance, granularity of group working by tutor and learners, and varied peer and tutor support, feedback, and discussion"* (Stiles, 2000). By contrast, in France, VLEs were conceived from the outset as a single workspace for both management and learning activities. The administrative modules (marks, absences) designed for virtual classrooms served then to design pedagogical applications and collaborative working groups. In both British and French systems, VLEs aim to encourage communication and collaborative practices between the members of a school community through tools like blogs or email and to foster access to information.

So we can see that VLEs serve to carry out diverse activities, are intended for several distinct user groups (teachers, students, parents, and staff), and can be exploited in very different contexts: in the classroom, at home or on the move. This complexity can limit the development of practices and the motivation to use it. In this article we chose to evaluate the factors involved in VLEs adoption in primary schools and to consider two processes: technology acceptance and appropriation. When they explain acceptance, the existing studies focus either on individual factors (like satisfaction, effort expectancy) or practical factors (technological features like ergonomic of the system), or, lately, contextual factors (like history and evolution of professional practices). In this article, we propose to present the main theoretical frameworks in the study of acceptance and to eventually describe a triangulated model to evaluate technology adoption. It represents a first version of a model of technology adoption that is based on different theoretical frameworks.

Most of the theoretical approaches which try to explain technology adoption are actually describing acceptance and appropriation and come from the fields of social psychology and ergonomics. This sections describe three positions: the model of social acceptance, practical acceptance and situated acceptance.

## 1.1 The Models of "Social Acceptance"

These approaches focus on human factors in the process of technological acceptance. The main idea is that people's perceptions and attitudes may play a major role in this process. According to Davis (1989) and his model TAM (Technology Acceptance Model), acceptance can be explained through two factors: perceived usefulness and perceived ease of use. These two perceptions influence the intentions to use the technology which, in turn, influence the acceptance of the technology. Other attitudinal factors are later added: satisfaction, performance expectancy, effort expectancy. This model is inspired by the theory of reasoned action (Ajzen, 1991) which consider that behaviour is guided from inside by people's intentions. Other authors (Blackwell et al., 2013) talk about internal factors (like beliefs, convictions and attitudes of users), and external factors (like support, training, technical infrastructure). Some authors support the idea that internal factors take priority in the decision to use an educational technology (Pynoo et al. 2011, Pynoo et al., 2012) while others think that external factors are predominant (Ertmer, 2005). When they study VLEs acceptance in particular, authors highlight the same duality. While some support the major role of technical infrastructure like access to the computer classroom, number of computers in classroom, Internet access and high speed Internet access and institution management (Keller, 2006, Keller, 2009, Osika, 2009, Babic, 2012), others admit that causes of VLEs reject are lack of confidence in technology and lack of time to train (Karasavvidis, 2009). Other studies show that it is actually the connection between the internal and the external factors that matters: external factor (like institutional support, training) will subsequently shape the beliefs and attitudes toward the technologies and then the intention to use those (Inan and Lowther, 2010).

In primary teaching, technology are less frequent, so there are not many studies on this particular subject. Studies demonstrated the importance of self-confidence toward computer use in the development of attitudes toward technologies and indirectly in the intention to use the technologies (Chen and Chang, 2006, Faurie and Van de Leemput, 2007, Giamalas and Nikolopulus, 2010, Tsytouridou and Vryzas, 2004). Beside confidence, some authors outline the role of perceived security in the acceptance of VLEs in primary school (Codreanu et al., 2015). VLEs suppose a functioning similar to that of social networks, with a unique access to content. Teachers doubt their own possibilities of control and moderation in cases of on-line bullying and interrogate about the responsibilities in case of misappropriation of the VLE by students. Also, they worry about the misuse identity by other colleagues. In primary schools, these issues are particular important, because the students are particular young and vulnerable to these forms of harassment.

Social acceptance approaches have nonetheless been subject to a number of criticisms concerning both methodological criteria and the models' foundations (Brangier, Dufresne, and Hammes-Adelé, 2009). One criticism is that these studies have little practical relevance for the technological design and improvement of the system. In effect, these studies indicate that a system is not acceptable to the target group without giving any information about the changes and adaptations required. Added to this is the fact that the research is based on small samples that are not representative of the professional context, and use questionnaires (scale of measurement) as the sole method of evaluation. Critics claim that such a method results in a truncated, partial and rather disembodied picture of the meaning people attach to the technology. However, in educational context, we retain the effort to specify precise factors directly implied in technological acceptance: confidence in computer use, social and institutional support, technological infrastructure and children's security.

## 1.2 The Models of "Practical Acceptance"

This approach focuses on the technology characteristics (human factors and ergonomics) and how the tool is implemented (support, training, participatory design). The prevailing idea is that when technology is easy to use and well implemented (training is provided and end users are included in the design process, for example) the device's acceptance is enhanced. In sum, the aim is not only to design a suitable product, but also a suitable relationship to technology, and ultimately contribute to an acceptable user experience for the individual (Barcenilla and Bastien, 2009).

According to Nielsen (1994), the two most important attributes for technology acceptance are usability and utility. Usability refers to the fact that people can easily use the functions of a system. Utility refers to the capacity of the system to help users do their tasks. In short, a technology easy to use and useful will be accepted by the users. To these two attributes, Nielsen adds others: costs of the technology, compatibility, reliability. We have to mention that the notion of "usability" is different of that of "perceived ease of use" in the previous social model. While the first refers to the effective usability and is evaluated through user tests, the second refers to perceptions and subjective attitude toward usability and is evaluated through questionnaires. The ISO 9241 norm specify that the three dimension of usability are: effectiveness (the accuracy with which users achieve specified goals), efficiency (the effort required for users to do theirs tasks) and satisfaction: what users think about the system.

Ergonomics specialists proposed a list of criteria to evaluate the usability of computer interfaces. Bastien and Scapin (1993) proposed eight criteria: guidance (means available to orient the user throughout the interface), workload (interface elements that play a role in the reduction of users' perceptual and cognitive load), explicit control (the control users have on the processing of their actions), adaptability (the system's capacity to behave according to users' needs), error management (means available to recover from errors), consistency (maintaining the interface choices in similar contexts), significance of codes (codes and names should be meaningful for users) and compatibility (match between the users characteristics and task characteristics). Concerning the last criteria, compatibility is particularly important when technologies are used by users with specific characteristics (in terms of age, customs, perceptions, skills). For instance, technologies designed to be used in primary schools, should be adapted to a public of young children, who do not master writing, reading and have limited fine motor skills. So, the interfaces should avoid using a lot of text content and complex pull-down menus; they should prefer instead images and simple menus (Hourcade, 2007, Lueder and Rice, 2008). Budiu and Nielsen (2010) used specific methods in order to evaluate children's behaviour on the web (think aloud, card sorting). They proposed a list of 130 recommendations for interfaces designed for children (aged 3 to 12), organised by the type of content (general interaction, navigation, images, videos etc.). Generally, they recommend to use interactive content, sound and colours, use of the metaphors and big buttons. They also advise to ensure children's control over the interface and to avoid sensory and cognitive overload.

These studies are important because they provide precious practical advising for designers. The main criticism is that they are focused on functional aspects and do not consider the intrinsic characteristics of user like emotions (pleasure, fun, amusement). Recently, studies began to consider user as a real partner in design of a technology in approaches like User Centred Design and participatory design (Carroll and Rosson, 2007, Carroll, 2008). Participatory design "relies on the collective generativity of stakeholders; in other words, it uses the collective ability of stakeholders to generate or create thoughts and imaginings" (Baek and Lee, 2008, pp. 173). In school technologies, participatory design suppose that teachers and students can be actively involved in the design of their future tools so that these tools would better meet their needs (Sucupura-Furtado, 2008, Konings et al., 2007, Konings, Seidel, and van Merrienboer 2014, Chin, 2004).

This approach focuses therefore on the technology conception, on ergonomic improvements and on support to collaboration between designers and end users. In this context, ergonomic approaches intend to prescribe recommendations and guidelines for designers in terms of technological adaption to users' needs. However, these studies remain focused on the functional aspects and on the performance of users with the system. In addition, participatory design, mostly applied in industry, is less adopted by the stakeholders in digital education. This is due, on one side, to the difficulty and high cost of putting participatory design into practice and, on the other side, to the diversity of educational contexts and high number of schools, with their own autonomy and specific organization which make technological generalization difficult.

## 1.3 Appropriation and Situated Acceptance Models

To address these limitations, the socio constructivist approaches (Engestrom, 1987, Engestrom, Mietinnen, and Punamaki, 1999) propose to take account of the modalities of use and features of the context in order to explain why and how a technology is accepted by users. The notion of appropriation is central. According to Engestrom (1987), a tool is not appropriated on neutral ground, but as part of a history of practices and a pre-existing culture. Engestrom proposes the notion of activity system, made up of a subject, a technological artefact, an object of activity, a community, operating rules, and a division of labour between the community members. In the school system, a new technology enters a context in which tools already exist – blackboards, pencils and textbooks, etc. – and have formatted how teachers work. The new tool may also alter the relationships between community members (teachers, students and parents). This confrontation between the new technology and the existing cultural-historical background can give rise to tensions or contradictions. These tensions favour and trigger innovation and change and are a source of development. The term "contradiction" should not be understood as a problem, barrier or conflict but in terms of development and progress.

According to Jonsson (2007), appropriation is "the gradual process by which participants successively become more proficient in using the tools" (p. 11) Unlike mastery, which entails the acquisition of a skill, appropriation, in addition to a technical skill, includes the competence to use the technology for carrying out an authentic task in a given context. As such, appropriation is thought to be strongly linked to the notion of change. Using a text editor at school changes practices very little, but being able to modify a digital text without having to copy it out can change the importance traditionally attached to writing.

Bobillier-Chaumon (2016) considers that the appropriation of a technological tool is a condition of its acceptance. When someone appropriates a tool, she contributes to it and is able to innovate, and therefore use the tool for previously unforeseen purposes. By making this contribution to the technology, the person can identify with it, make it her own, give it meaning and therefore accept it. Bobillier-Chaumon proposes the notion of situated acceptance, defined "as the way in which an individual – or a group or organization – perceives the issues related to these technologies (strengths, benefits, risks, opportunity) through their use in everyday situations, and reacts to them (favourably or not)." (Bobillier-Chaumon and Dubois, 2009). What is taken into account here is the experience in a situation of interaction between users and a certain technology that already exists. In this approach, the object of study is not the perception or attitude towards technology but the practices and activities carried out as part of a real job.

The advantage of this approach is that it brings into light for the first time dimensions like "history" and "context" and proposes to look for acceptance directly in daily activities of end users. Methodology consists of qualitative studies (case studies, activity analysis, and elicitation interviews), small samplings and a certain "opening" of the researchers: they do not depend on *a priori* hypothesis. Their recommendations are highly adapted to the situation and identify issues that are not previously visible or expected. The main criticism rely on the fact that situated acceptance models focus on specific situations and it may be difficult to replicate them in other contexts. Therefore, we propose in this article a prospective model to evaluate acceptance, which could be used in different educational contexts.

## 2 AN ANALITICAL MODEL OF VLE ADOPTION IN PRIMARY SCHOOLS

We have identified three categories of approaches. The first, social acceptance, focuses on the individual perceptions and attitudes of prospective users; the second, practical acceptance, concentrates on the tool's ergonomic characteristics; and the third analyses users' activities and hence the interaction between the technology and actual practices. In our study, we need to evaluate the acceptance of a VLE, a complex tool designed for multiple user groups (teachers, students and parents) to perform diverse tasks in a range of contexts (communication, learning, monitoring, etc.). Consequently, we consider that acceptance is a process that can be evaluated through three sets of factors:

- Technological factors grouped in system quality factors (like usability) and design quality factors (participatory design)
- Activity and task factors related to characteristics of professional activity like rules, prescriptions, professional practices, objectives

- Perception factors related to individual opinions about the qualities of the technology (perceived ease of use, satisfaction, perceived security)

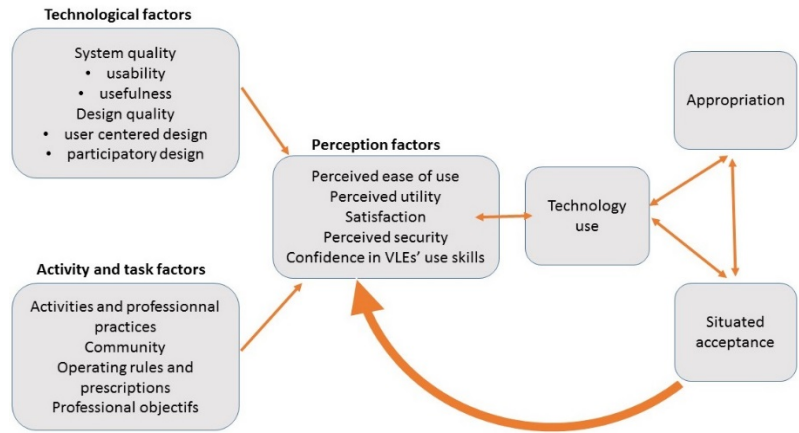

Figure 1. Model of VLE adoption: factors involved in the acceptance and appropriation of VLEs.

In the above diagram (Figure 1), the single arrow indicates a one-way relationship between the two factor categories. The double arrow indicates a two-way relationship. The technological factors (quality of the product, quality of support) influence the perceptions of the tool, which in turn influence the tool's appropriation and acceptance. For their part, the activity and task factors (activity, practices, community) also influence the perception factors. The creation of technology's meaning is made during the actual use. The use trials influence significantly the level of technology acceptation and appropriation. The quality of use will build a new form of appropriation (by creating new forms of practices and innovative use) and acceptance (through the lens of new emotions and new benefits related to use). These two constructs will modify the initial perception of the technology and the users' perceptions on their technological skills. The retroactive loop describes how appropriation (seen as mastery of the tool plus innovation) is decisive for the acceptance of the tool (seen as the subjective decision to start using the technology) and vice versa.

It is a dynamic model that may enable the plurality of viewpoints and situations to be reconstructed. Dynamism of the model is important for explaining the principles of technology adoption through articulation of factors issued of different theoretical approaches. This model may restore a diversity of points of view and situations and the formalisation of factors' progression in context. In order to deepen this approach and qualify the criteria of each factor, we propose to use triangulated methods (Denzin, 1978) which consists in using more than one method to study a phenomenon. So, our model is based on a theory triangulation (using more than one theoretical scheme to interpret a phenomenon) and a methodological triangulation (using more than one method to study a phenomenon). In terms of methodology, we propose a triangulation consisting of qualitative methods (interviews, elicitation interviews, content analysis) and quantitative methods (questionnaires, analysis of connection logs).

We intend to illustrate this model in a new study. This research will include three different approaches: 1. an evaluation of the platform's ergonomics through user tests; 2. an evaluation of teachers and parents' perceptions about the VLE through questionnaires; and 3. an analysis of activities realised on the VLE by teachers, students and parents through thematic analysis of contributions made on VLE and interviews. In the first approach we intend to see if the VLE used is easy to use and adapt to the public, especially the young children. The other objective is to produce recommendations to designers in order to ameliorate the solution if needed. The second approach aims at collecting users' opinions about the VLE, on different criteria: perceived ease of use, satisfaction, usefulness in theirs activities, perceived security. The third approach consists of analysing real activities realised by teachers, students and parents with the VLE. The objective is to see how exactly they adapt the technology to their practices, and on what type of activities appropriation is constructed. For instance, we are interested to know if the teachers prefer using the VLE in order to provide communication and collaboration with parents or to realise pedagogical tasks with students. In Figure 2 we can see an example of a pedagogical contribution of a 7 years aged student, made on the VLE ONE. These answers may help us know what the priorities of users are and how they relate to the technology when it is first introduce, what use they represent in first and what activity they experiment. The advantage of this kind of model is that it proposes a large exploration of the subject of acceptance, through different angles of research and complementary methods (qualitative and quantitative). Another advantage is the fact that it may restore multiple points of view of educational community members: teachers, students and parents and the relation between these members. It permits a focus on different actors and their specific needs and characteristics.

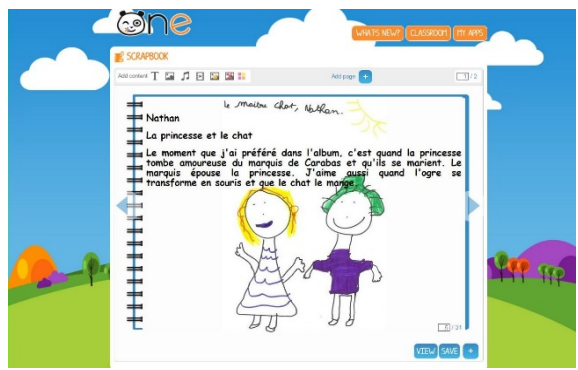

Figure.2. Example of pedagogical use of the Multimedia Notebook on VLE ONE.

## 3 CONCLUSIONS

In this paper we presented three important models in the study of technological adoption. The three models have their origins in different fields of research. The models of "social acceptance", like TAM and UTAUT were inspired by social psychology but applied to management and marketing studies. The "practical acceptance" theories are specific to ergonomists and designers. And finally, models of "situated acceptance" are also issued from development psychology and lately applied to various fields, from change management to organization issues. The technological adoption issue is of general interest and should not be limited to one singular approach. Our objective was to resume these various models and to extract information that is salient for educational area. Factors like usability for young children, teachers' confidence in their computer use skills, teacher's perceived security toward children's use and preexistent teaching practices are example of important determinants of technology adoption in schools. The proposed model represent a first theoretical proposition and it will subsequently be validated in specific studies.

## ACKNOWLEDGEMENTS

This paper has been conducted in the context of a Phd thesis which was funded by the company WebServices pour l'Education, Paris, France and supported by ANRT France (Association Nationale de la Recherche et de la Technologie). The VLE studied is called ONE (http://one1d.fr/en/home-page/) and it was specifically designed for an elementary school audience, with ergonomics and interfaces that are suitable for children. It is currently used by primary schools in France and South America.